\documentclass[seceq]{ptptex}

\usepackage{graphicx}

\pubinfo{January 2010}
\notypesetlogo                        

\title{
Characteristic Scales of Initial Density and Velocity Fields
}

\author{
Hideaki \textsc{Mouri}$^1$
and
Yoshiaki \textsc{Taniguchi}$^2$
}

\inst{
$^1$Meteorological Research Institute, Tsukuba 305-0052, Japan
\\
$^2$Research Center for Space and Cosmic Evolution, Ehime University,\\ Matsuyama 790-8577, Japan
}

\recdate{August 29, 2009}

\abst{
For the initial fields of the density contrast and peculiar velocity, we theoretically calculate the `differential' and `integral' length scales, i.e., statistical measures that respectively characterize the small- and large-scale fluctuations of a random field. These length scales and the associated mass scales explain the length and mass scales observed for (1) halos of young galaxies at $z \gtrsim 5$, (2) halos of galaxies at $z \simeq 0$, and (3) the largest structures in the galaxy distribution at $z \simeq 0$. We thereby discuss that such observed scales are fossil imprints of the characteristic scales of the initial fields.
}

\PTPindex{464, 469}

\begin{document}

\maketitle

\section{Introduction} \label{s1}

Since structures of the Universe are formed from gravitational instability of the initial fluctuations, some characteristics of the structures are traced back to those of the initial fluctuations. This should be especially the case for the characteristics determined in the linear or quasi-linear stage of the structure formation. Both the stages are described with the initial peculiar velocity {\boldmath $v$} by the first-order Lagrangian perturbation.\cite{Ze70,SC95} The linear stage is also described with the initial density contrast $\delta$ by the first-order Eulerian perturbation.\cite{SC95,Pe93}

The characteristics of the initial fields of the density contrast $\delta$ and peculiar velocity {\boldmath $v$} are studied here. Since these fields are random, we use statistical measures that have been developed to characterize turbulent random fields.\cite{MY75} Specifically, for the first time, the `differential' and `integral' length scales are used to characterize the initial fields. We calculate the length scales and associated mass scales. They are compared with the scales observed for galaxies and their large-scale distribution, in order to explore the fossil imprints of the initial fluctuations among these structures.

\section{Basic framework}
\label{s2}
\subsection{Cosmological parameters}
\label{s21}

For the cosmological parameters, we assume their standard values: the Hubble constant $H_0 = 70$\,km\,s$^{-1}$\,Mpc$^{-1}$ ($h = 0.7$), the matter density $\Omega_{\rm m} = 0.14 h^{-2}$, the baryon density $\Omega_{\rm b} = 0.023 h^{-2}$, and the temperature of the cosmic microwave background $T_{\rm CMB} = 2.73$\,K.\cite{Ko08} The matter density $\Omega_{\rm m}$ is dominated by the cold dark matter so that the structure formation proceeds from small to large scales.\\

\subsection{Density and velocity fluctuations}
\label{s22}

The density contrast $\delta (\mbox{\boldmath $x$},t)$ at comoving position {\boldmath $x$} and at time $t$ is defined as $\delta (\mbox{\boldmath $x$},t) = \rho (\mbox{\boldmath $x$},t) / \langle \rho \rangle -1$ with the mass density $\rho (\mbox{\boldmath $x$},t)$. Here, $\langle \cdot \rangle$ denotes an average and $\langle \rho \rangle = 3 \Omega _{\rm m} H_0^2 / 8 \pi G$. During the linear stage, the linear growth factor $D(t)$ describes the evolution as $\delta (\mbox{\boldmath $x$},t) = D(t) \delta (\mbox{\boldmath $x$})$.

The density contrast $\delta (\mbox{\boldmath $x$})$ at the initial time $t_{\rm in}$ makes up a random field, which is related to its Fourier transform $\tilde{\delta}(\mbox{\boldmath $k$})$ as
\begin{equation}
\label{e1}
\delta (\mbox{\boldmath $x$}) 
=
\int \tilde{\delta} (\mbox{\boldmath $k$}) 
     \exp ( i \mbox{\boldmath $k$} \cdot \mbox{\boldmath $x$})  
\frac{d \mbox{\boldmath $k$}}{(2 \pi)^{3/2}}.
\end{equation}
Since $\delta (\mbox{\boldmath $x$})$ is homogeneous, isotropic, and Gaussian, at least as a good approximation, its statistics are determined by the initial power spectrum $P_{\delta}(k)$ for wave number $k = \vert \mbox{\boldmath $k$} \vert$:
\begin{equation}
\label{e2}
\langle \tilde{\delta} (\mbox{\boldmath $k$} ) 
        \tilde{\delta} (\mbox{\boldmath $k$}')^{\ast} \rangle
=
(2 \pi )^3 \delta_{\rm D} (\mbox{\boldmath $k$} - \mbox{\boldmath $k$}') P_{\delta}(k). 
\end{equation}
Here, $\tilde{\delta} (\mbox{\boldmath $k$})^{\ast}$ is the complex conjugate of $\tilde{\delta} (\mbox{\boldmath $k$} )$, and $\delta_{\rm D}(\mbox{\boldmath $k$})$ is Dirac's delta function.

For the growing mode of the fluctuations, the peculiar velocity $\mbox{\boldmath $v$}(\mbox{\boldmath $x$})$ at $t = t_{\rm in}$ is defined by using $\tilde{\delta}(\mbox{\boldmath $k$})$:
\begin{equation}
\label{e3}
\mbox{\boldmath $v$}(\mbox{\boldmath $x$})
=
\dot{D}(t_{\rm in})
\int  \frac{ i \mbox{\boldmath $k$} \tilde{\delta}(\mbox{\boldmath $k$}) }{k^2}            
      \exp ( i \mbox{\boldmath $k$} \cdot \mbox{\boldmath $x$}) 
\frac{d \mbox{\boldmath $k$}}{(2 \pi)^{3/2}}
\end{equation}
(Peebles\cite{Pe93}, p.~514), where $\dot{D}(t)$ is the time derivative of $D(t)$. The factor $\mbox{\boldmath $k$} /k^2$ implies that $\mbox{\boldmath $v$} (\mbox{\boldmath $x$})$ is dominated by larger scales than $\delta (\mbox{\boldmath $x$})$. Length scales for $\mbox{\boldmath $v$} (\mbox{\boldmath $x$})$ are larger than those for $\delta (\mbox{\boldmath $x$})$ (see \S \ref{s4}).

The peculiar gravitational potential $\phi (\mbox{\boldmath $x$})$ at $t = t_{\rm in}$ is also defined by using $\tilde{\delta}(\mbox{\boldmath $k$})$:
\begin{equation}
\label{e4}
\phi (\mbox{\boldmath $x$}) 
=
- 
\frac{4 \pi G \langle \rho \rangle}{\left[ 1+z(t_{\rm in}) \right]^2}
\int \frac{\tilde{\delta} (\mbox{\boldmath $k$})}{k^2}     
     \exp ( i \mbox{\boldmath $k$} \cdot \mbox{\boldmath $x$}) 
\frac{d \mbox{\boldmath $k$}}{(2 \pi)^{3/2}}
\end{equation}
(Peebles\cite{Pe93}, p. 114), where $z(t)$ is the redshift. We have $\mbox{\boldmath $v$}(\mbox{\boldmath $x$}) \propto -\mbox{\boldmath $\nabla$} \phi(\mbox{\boldmath $x$})$. That is, at $t = t_{\rm in}$, the peculiar velocity $\mbox{\boldmath $v$}(\mbox{\boldmath $x$})$ for the growing mode is proportional to the peculiar gravitational acceleration $-\mbox{\boldmath $\nabla$} \phi(\mbox{\boldmath $x$})$.

Usually, the initial fields are smoothed with a window function.\cite{SC95,Pe93} The reason is the significant small-scale fluctuations retained by the cold dark matter. For example, Mouri and Taniguchi\cite{MT05} smoothed $\delta (\mbox{\boldmath $x$})$ and $\mbox{\boldmath $v$}(\mbox{\boldmath $x$})$ to calculate some of their characteristic scales. However, any smoothing imposes its scale as an additional artificial scale. The present study is to be among the first to calculate scales of the unsmoothed initial fields.

\subsection{Perturbation theories}
\label{s23}

The first-order Eulerian perturbation $\delta (\mbox{\boldmath $x$},t) = D(t) \delta (\mbox{\boldmath $x$})$ has been mentioned in \S \ref{s22}. This holds for regions in the linear stage. The absolute value of $\delta (\mbox{\boldmath $x$},t)$ averaged over such a region is $\ll 1$.

The first-order Lagrangian perturbation, i.e., the Zel'dovich approximation,\cite{Ze70} describes the position $\mbox{\boldmath $x$}(t)$ of a matter element with its initial position {\boldmath $x$}$_{\rm in}$ and its initial peculiar velocity {\boldmath $v$}$(${\boldmath $x$}$_{\rm in})$: 
\begin{equation}
\label{e5}
\mbox{\boldmath $x$}(t) 
= 
\mbox{\boldmath $x$}_{\rm in} + \frac{D(t)}{\dot{D}(t_{\rm in})}\mbox{\boldmath $v$}(\mbox{\boldmath $x$}_{\rm in}).
\end{equation}
This holds for regions in the linear or quasi-linear stage. The absolute value of $\delta (\mbox{\boldmath $x$},t)$ averaged over such a region is $\lesssim 1$. In other words, Eq. (\ref{e5}) is valid so far as the matter elements from different initial positions do not come across one another.\cite{SC95,Yo06}

\subsection{Models for structure formation}
\label{s24}

The perturbation theories allow us to relate scales of the initial fields to scales of structures determined in the linear or quasi-linear stage of their formation. However, these structures eventually become nonlinear. To such self-gravitating objects with $\delta (\mbox{\boldmath $x$},t) \gg 1$, the perturbed initial fields are extended by using models.

For scales of the density field, we use the peak model of Kaiser.\cite{SC95,Pe93,Ka84} This is based on the Eulerian perturbation $\delta (\mbox{\boldmath $x$},t) = D(t) \delta (\mbox{\boldmath $x$})$. When $D(t) \delta (\mbox{\boldmath $x$})$ averaged over a region exceeds some critical value of order unity, the region is assumed to collapse and form a self-gravitating object. It follows that the structure formation occurs preferentially around peaks of the initial density field.

For scales of the velocity field, we use the adhesion model of Gurbatov et al.\cite{SC95,Gu89} This is based on the Lagrangian perturbation in Eq. (\ref{e5}). When matter elements from different initial positions come across one another, the adhesion model assumes that the matter elements adhere together, mimicking the gravitational interaction and the resultant formation of a self-gravitating object.

Thus, for scales of the density and velocity fields, we use different models, although they represent the same dynamics. Details of the models are not important here because we only require their qualitative predictions.

\section{Characteristic scales}
\label{s3}
\subsection{Definition and meaning}
\label{s31}

Consider a homogeneous isotropic random field $f(\mbox{\boldmath $x$})$ with $\langle f(\mbox{\boldmath $x$}) \rangle = 0$. Its two-point correlation at scale $r = \vert \mbox{\boldmath $r$} \vert$ is $\langle f(\mbox{\boldmath $x$})f(\mbox{\boldmath $x$}+\mbox{\boldmath $r$}) \rangle$, which is used to define the differential and integral length scales. Since these length scales possess not only the dimensions of length but also particular meanings, they deserve to be compared with the observed scales.

The differential length scale $\lambda$ is defined by using the correlation and its second-order derivative at $r = 0$ as
\begin{equation}
\label{e6}
\lambda
=
\pi \left[ - \frac{\langle f(\mbox{\boldmath $x$})^2 \rangle}
       {\partial_r^2 \langle f(\mbox{\boldmath $x$})
                             f(\mbox{\boldmath $x$}+\mbox{\boldmath $r$}) \rangle \vert_{r=0}} 
    \right] ^{1/2}.
\end{equation}
This is identical to $\lambda = \pi \left[ \langle f(x)^2 \rangle / \langle (\partial_x f)^2 \rangle  \right] ^{1/2}$, where $x$ is any one-dimensional cut of the space {\boldmath $x$} (Monin and Yaglom,\cite{MY75} pp. 25 and 35). The original definition lacks the factor $\pi$. Our definition with the factor $\pi$ is essential so that $\lambda$ has a particular meaning. That is, if $f(\mbox{\boldmath $x$})$ is Gaussian as in the case of the initial density and velocity fields, $\lambda$ is exactly identical to the mean interval between successive zero crossings of $f(\mbox{\boldmath $x$})$ along any one-dimensional cut of the space.\cite{Ri54} Although past studies used scales that differ from $\lambda$ only by numerical factors,\cite{SC95,Ba86,Ko92} the difference is essential so that $\lambda$ is a new statistical measure with the new particular meaning.

The integral length scale $L$ is defined by integrating the correlation over all $r$ as
\begin{equation}
\label{e7}
L = \frac{\int ^{\infty}_0 \langle f(\mbox{\boldmath $x$})f(\mbox{\boldmath $x$}+\mbox{\boldmath $r$}) \rangle dr}
         {\langle f(\mbox{\boldmath $x$})^2 \rangle}
\end{equation}
(Monin and Yaglom,\cite{MY75} p. 34). This is the characteristic length scale of significant correlation. While $\langle f(\mbox{\boldmath $x$})f(\mbox{\boldmath $x$}+\mbox{\boldmath $r$}) \rangle / \langle f(\mbox{\boldmath $x$})^2 \rangle$ is significant at $r \lesssim L$, it is not significant at $r \gtrsim L$. An elementary example is a correlation that is proportional to an exponential function $\exp (-r/L)$ or a step function $\theta (L-r)$. The situation is the same for correlations in the initial density and velocity fields (see \S \ref{s4}). To characterize such correlations, past studies used scales $\propto \lambda$.\cite{SC95,Ko92}. However, they are not the length scales of significant correlation because $\lambda$ is defined only at $r = 0$ (Eq. (\ref{e6})). In fact, we usually have $\lambda \ll L$.

\subsection{Condition for existence}
\label{s32}

There is a condition required for the existence of the length scales $\lambda$ and $L$. They have been defined with the correlation function, which is in turn defined with the power spectrum. For the initial density and velocity fields studied here, the length scales are in the forms of
\begin{equation}
\label{e8}
\lambda 
\propto 
\left[ \frac{\int ^{\infty}_0 k^{m} P_{\delta}(k) dk}{\int ^{\infty}_0 k^{m+2} P_{\delta}(k) dk} \right]^{1/2}
\quad
\mbox{and}
\quad
L
\propto
\frac{\int ^{\infty}_0 k^{m-1} P_{\delta}(k) dk}{\int ^{\infty}_0 k^{m} P_{\delta}(k) dk},
\end{equation}
with integer $m$ (Eqs. (\ref{e12}), (\ref{e14}), (\ref{e16}), and (\ref{e17})). The above integrals are required to converge. This condition is important especially because we use the unsmoothed initial fields (\S \ref{s22}).

To describe the initial power spectrum, we have $P_{\delta}(k) \propto T(k)^2 k^{n_s}$. Since the transfer function $T(k)$ is dominated by the cold dark matter, it has the asymptotes\cite{Ba86}
\begin{equation}
\label{e9}
\lim_{k \rightarrow 0}      T(k) =       1              \quad \mbox{and} \quad
\lim_{k \rightarrow \infty} T(k) \propto \frac{\ln (k)}{k^2}. 
\end{equation}
Here, we have ignored the cutoff at a very high $k$ that reflects the particle physics of the cold dark matter,\cite{Gr04,Pr06} for which little is known. The hot dark matter and baryon are not dominant and hence affect the asymptotes only through the prefactor of $\ln (k)/k^2$.\cite{EH98,HE98} On the other hand, for the primordial power spectrum $k^{n_s}$, the standard model is the Harrison-Zel'dovich spectrum with $n_s = 1$. However, the actual spectrum for the observable Universe is tilted:\cite{Ko08}
\begin{equation}
\label{e10}
           n_s = 1-\Delta_s \quad \mbox{with} \quad \Delta_s \simeq 0.04,
\end{equation}
because of some deviation from exact exponential expansion of the Universe during the corresponding stage of the inflation.

Hence, in Eq. (\ref{e8}), while the integrals for $\lambda$ converge if $m$ is $-1$ or 0, those for $L$ converge if $m$ is 0, 1, or 2. This condition is satisfied by the density integral scale in Eq. (\ref{e12}), velocity differential scale in Eq. (\ref{e16}), and velocity integral scale in Eq. (\ref{e17}), but not by the density differential scale in Eq. (\ref{e14}). The last scale does not exist.\footnote{
If we assume the Harrison-Zel'dovich model spectrum with $n_s = 1$, only the velocity integral scale exists. For the other scales, the integrals do not converge because small-scale fluctuations of the unsmoothed initial fields are too significant (see also \S \ref{s22}). Thus, it is essential to use the actual spectrum with $n_s = 1-\Delta_s \simeq 0.96$.}
 It should also be noted that the condition is not satisfied by either $\lambda$ or $L$ for the initial field of gravitational potential.

\begin{table}[tbp]
\begin{center}
\begin{minipage}{8.6cm}
\begin{center}

\caption{Length scales $L$ and $\lambda$ and associated mass scales $M(L) = 4 \pi \langle \rho \rangle L^3/3$ and $M(\lambda) = 4 \pi \langle \rho \rangle \lambda^3/3$ for $n_s = 0.96 \pm 0.01$, $\Omega_{\rm m} h^2 = 0.14$, $\Omega_{\rm b} h^2 = 0.023$, and $T_{\rm CMB} = 2.73$\,K. The scales are smaller for the greater value of $n_s$.}
\label{t1}

\begin{tabular}{@{ }llcllc@{ }}
\hline \hline
\noalign{\smallskip}
\multicolumn{2}{c}{Length}                              && \multicolumn{2}{c}{Mass}                                              & Eq.    \\
\cline{1-2}                                                \cline{4-5} 
\noalign{\smallskip}
                                & \ \ (Mpc)             &&                                     & \quad \ \ $(M_{\odot})$         &             \\
\hline 
\noalign{\smallskip}
$L_{\delta}$                    &$0.023^{-0.004}_{+0.004}$&&$M(L_{\delta})$                    &$2.0^{-0.8}_{+1.2}\times 10^6$   &(\ref{e12}) \\ 
\noalign{\smallskip}
$\lambda_{v_{\parallel}}$       &$1.5  ^{-0.2}  _{+0.1}$  &&$M(\lambda_{v_{\parallel}})$       &$5.0^{-1.3}_{+1.7}\times 10^{11}$&(\ref{e16a})\\
\noalign{\smallskip}
$\lambda_{v_{\perp}}$           &$2.5  ^{-0.2}  _{+0.3}$  && $M(\lambda_{v_{\perp}})$          &$2.6^{-0.7}_{+0.9}\times 10^{12}$&(\ref{eA2}) \\
\noalign{\smallskip}
$\lambda_{\mbox{\boldmath $v$}}$&$2.0  ^{-0.2}  _{+0.1}$  &&$M(\lambda_{\mbox{\boldmath $v$}})$&$1.2^{-0.3}_{+0.4}\times 10^{12}$&(\ref{e16b})\\
\noalign{\smallskip}
$L_{v_{\perp}}$                 &$130. ^{-2.}   _{+2.}$   && $M(L_{v_{\perp}})$                &$3.6^{-0.2}_{+0.2}\times 10^{17}$&(\ref{eA3}) \\
\noalign{\smallskip}
$L_{\mbox{\boldmath $v$}}$      &$87.  ^{-2.}   _{+1.}$   && $M(L_{\mbox{\boldmath $v$}})$     &$1.1^{-0.1}_{+0.0}\times 10^{17}$&(\ref{e17a})\\ 
\noalign{\smallskip}
\hline
\end{tabular}

\end{center}
\end{minipage}
\end{center}
\end{table}

\section{Results and discussion}
\label{s4}

Table \ref{t1} lists the numerical values of the length scales $L$ and $\lambda$ as well as of the associated mass scales $M(L) = 4 \pi \langle \rho \rangle L^3/3$ and $M(\lambda) = 4 \pi \langle \rho \rangle \lambda^3/3$.

For the primordial power spectrum $k^{n_s}$, since the $n_s$ value is crucial to the convergence of the integrals in Eq. (\ref{e8}), we have used the value with an uncertainty, $n_s = 0.96 \pm 0.01$.\cite{Ko08} Table \ref{t1} shows that the corresponding uncertainties are relatively less significant in the larger scales.

For the transfer function $T(k)$, we have used the fitting formulae of Eisenstein and Hu.\cite{EH98} Their input parameters are $\Omega_{\rm m} h^2$, $\Omega_{\rm b} h^2$, and $T_{\rm CMB}$, for which we have used the standard values in \S \ref{s21}.\footnote{
If we assume an uncertainty of $\pm 0.01$ in $\Omega_m h^2$, $\pm 0.001$ in $\Omega_b h^2$, or $\pm 0.01$\,K in $T_{\rm CMB}$, it leads to an uncertainty of about $\mp 20$\%, $\pm 3$\%, or $\pm 2$\% in each of the mass scales.} 
 Since $\Omega_{\rm b} h^2 \ll \Omega_{\rm m} h^2$, features of the baryon acoustic oscillations\cite{EH98} are not so strong in $T(k)$. Then, the overall shape of $P_{\delta}(k) \propto T(k)^2 k^{n_s}$ determines the values of the scales. The hot dark matter, i.e., neutrino, has been  safely ignored because its particle mass, albeit not exactly known, is small enough.\cite{Ko08}

Hereafter, the individual scales are studied along the structure formation from small to large scales, or equivalently, from early to late times.

\begin{figure}[tbp]
\begin{center}
\resizebox{8.4cm}{!}{\includegraphics*[2.9cm,13.2cm][17.7cm,25cm]{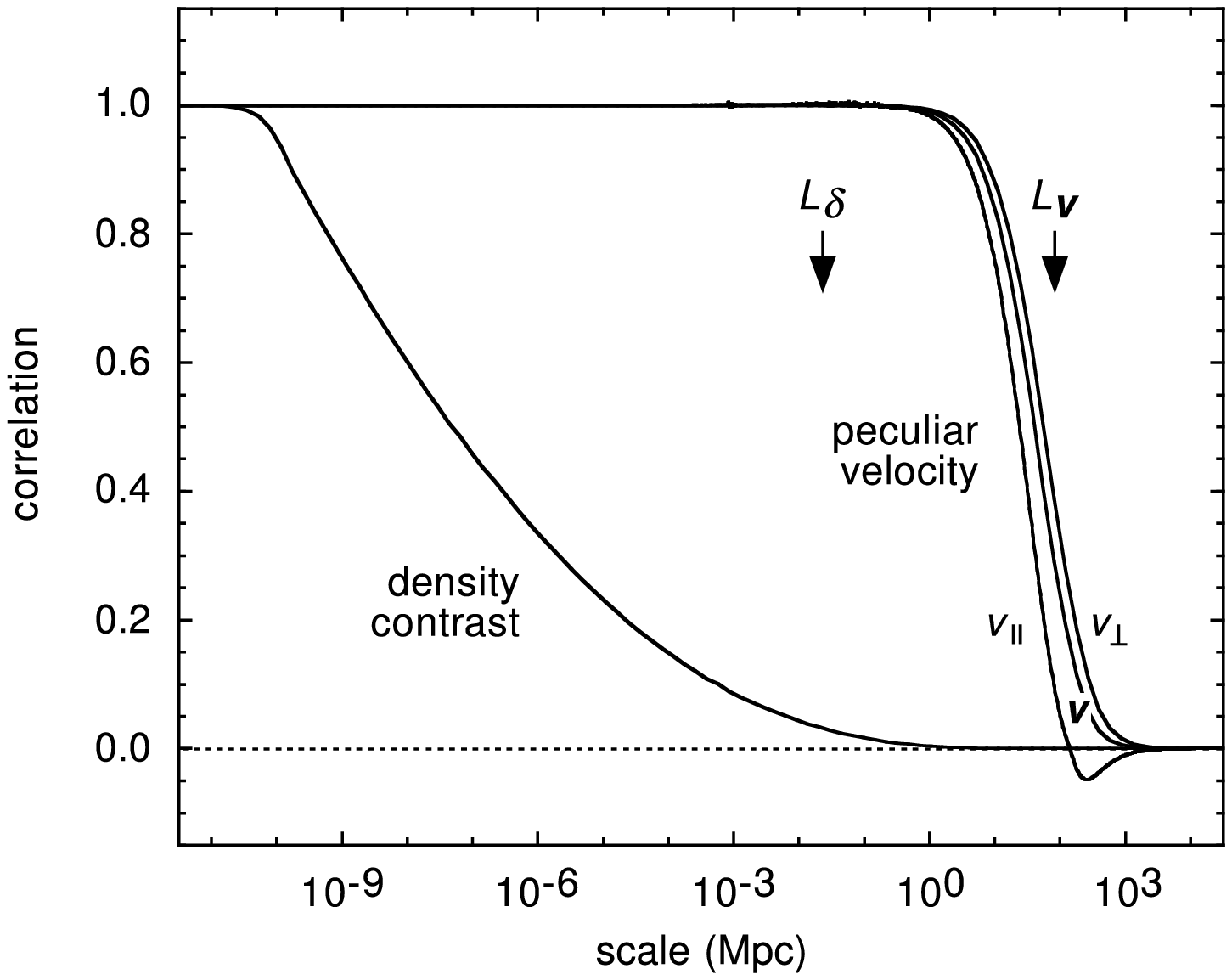}}

\caption[Fig1]{Two-point correlations $\langle \delta (\mbox{\boldmath $x$}) \delta (\mbox{\boldmath $x$} + \mbox{\boldmath $r$}) \rangle$, $\langle \mbox{\boldmath $v$} (\mbox{\boldmath $x$}) \cdot \mbox{\boldmath $v$} (\mbox{\boldmath $x$} + \mbox{\boldmath $r$}) \rangle$, $\langle v_{\parallel} (\mbox{\boldmath $x$}) v_{\parallel} (\mbox{\boldmath $x$} + \mbox{\boldmath $r$}) \rangle$, and $\langle v_{\perp} (\mbox{\boldmath $x$}) v_{\perp} (\mbox{\boldmath $x$} + \mbox{\boldmath $r$}) \rangle$ at the initial time $t_{\rm in}$ as a function of scale $r$. They are normalized by the values at $r = 0$. The arrows indicate the integral length scales $L_{\delta}$ and $L_{\mbox{\boldmath $v$}}$.
\label{f1}}

\end{center}
\end{figure}

\subsection{Density integral scale}
\label{s41}

By using Eqs. (\ref{e1}) and (\ref{e2}), from the initial power spectrum $P_{\delta}(k)$, we obtain the two-point correlation of the initial density contrast $\delta (\mbox{\boldmath $x$})$ as
\begin{equation}
\label{e11}
\langle \delta (\mbox{\boldmath $x$}) \delta (\mbox{\boldmath $x$} + \mbox{\boldmath $r$}) \rangle
=
4 \pi \int ^{\infty}_{0} k^2 j_0 (kr) P_{\delta}(k) dk 
\end{equation}
(Peebles,\cite{Pe93} p. 509), where $j_0(x) = \sin (x)/x$ is a spherical Bessel function. The profile of $\langle \delta (\mbox{\boldmath $x$}) \delta (\mbox{\boldmath $x$} + \mbox{\boldmath $r$}) \rangle/ \langle \delta (\mbox{\boldmath $x$}) ^2 \rangle$ is shown in Fig. \ref{f1}.

Equations (\ref{e7}) and (\ref{e11}) are used to obtain the integral length scale for the initial density contrast:
\begin{equation}
\label{e12}
L_{\delta}
=     
\frac{\pi \int ^{\infty}_0 k P_{\delta}(k) dk}{2 \int ^{\infty}_0 k^2 P_{\delta}(k) dk}.
\end{equation}
Table \ref{t1} shows that the length scale $L_{\delta}$ is about 0.02\,Mpc and the associated mass scale $M(L_{\delta})$ is about $2 \times 10^6$\,$M_{\odot}$. We regard them as lower limits. The actual values could be much greater. This is because we have ignored the high-$k$ cutoff in $P_{\delta}(k)$ (\S \ref{s32}). To the other scales in Table \ref{t1}, the cutoff is not important.\footnote{
If we assume the cutoff $\exp[-(k/k_{\rm c})^2]$ at a relatively low wave number $k_{\rm c} = 10^6$\,Mpc$^{-1}$,\cite{Gr04,Pr06} $M(L_{\delta})$ is 37 times greater, while $M(\lambda_v)$ and $M(\lambda_{\mbox{\boldmath $v$}})$ are 6.1 times greater. If $k_{\rm c} = 10^7$ or $10^8$\,Mpc$^{-1}$, $M(L_{\delta})$ is 14 or 5.8 times greater, while $M(\lambda_v)$ and $M(\lambda_{\mbox{\boldmath $v$}})$ are 3.7 or 2.4 times greater. For each $k_{\rm c}$, $M(L_v)$ and $M(L_{\mbox{\boldmath $v$}})$ remain the same.}

The mass scale $M(L_{\delta})$ serves as the typical mass for overdense regions around high density peaks of the initial fluctuations. If such a peak at $\mbox{\boldmath $x$}_{\rm p}$ has the density contrast $\delta (\mbox{\boldmath $x$}_{\rm p})$, the mean density contrast at distance $r$ is  
\begin{equation}
\label{e13}
\langle \delta (\mbox{\boldmath $x$}_{\rm p}+\mbox{\boldmath $r$}) \rangle
=
\frac{\langle \delta (\mbox{\boldmath $x$}) \delta (\mbox{\boldmath $x$} + \mbox{\boldmath $r$}) \rangle}
     {\langle \delta (\mbox{\boldmath $x$}) ^2 \rangle}
\delta (\mbox{\boldmath $x$}_{\rm p})
\end{equation}
(Kaiser,\cite{Ka84} see also Peebles,\cite{Pe93} p. 623). This is identical to the radial profile of the mean overdensity. Figure \ref{f1} shows that $\langle \delta (\mbox{\boldmath $x$}) \delta (\mbox{\boldmath $x$} + \mbox{\boldmath $r$}) \rangle/ \langle \delta (\mbox{\boldmath $x$}) ^2 \rangle$ is significant up to $r \simeq L_{\delta}$ (\S \ref{s31}). The mean overdensity $\langle \delta (\mbox{\boldmath $x$}_{\rm p}+\mbox{\boldmath $r$}) \rangle$ is thereby significant up to the radius of $r \simeq L_{\delta}$. Then, $L_{\delta}$ serves as the typical radius for the overdense regions, and hence $M(L_{\delta}) = 4 \pi \langle \rho \rangle L_{\delta}^3 /3$ serves as the typical mass.

In practice, any overdense region consists of many density peaks. Their typical size is just the differential length scale for $\delta (\mbox{\boldmath $x$})$, i.e., mean interval between its zeros along any one-dimensional cut of the space (\S \ref{s31}):
\begin{equation}
\label{e14}
\lambda _{\delta}
=
\pi \left[ \frac{ 3 \int ^{\infty}_0 k^2 P_{\delta}(k) dk}{ \int ^{\infty}_0 k^4 P_{\delta}(k) dk} \right] ^{1/2}.
\end{equation}
Since the integral $\int ^{\infty}_0 k^4 P_{\delta}(k) dk$ diverges, the scale $\lambda _{\delta}$ is always zero. The sizes of those density peaks are very small.

The overdense region around a density peak $\delta (\mbox{\boldmath $x$}_{\rm p})$ collapses and forms a self-gravitating object, according to the peak model in \S \ref{s24}. If $\delta (\mbox{\boldmath $x$}_{\rm p})$ is high enough, the collapse occurs early enough so as not to be affected by other overdense regions. The object mass is typically $M(L_{\delta})$ that has a value at $> 2 \times 10^6$\,$M_{\odot}$. Being more massive than `minihalos' with $10^5$--$10^6$\,$M_{\odot}$ of the first stars,\cite{Yo03,Br09} this class of objects formed at early $t$ is considered to be a young stage of galactic halos once existent at high $z$.

The representative young galaxies observed at $z \gtrsim 5$ are `Ly$\alpha$ emitters' that are characterized by intense Ly$\alpha$ emission from ionized gas around hot massive stars. For the individual Ly$\alpha$ emitters observed so far, the range of the stellar mass is down to $10^6$\,$M_{\odot}$.\cite{El01,Pi07} This is consistent with the total halo mass of $M(L_{\delta})$ at $> 2 \times 10^6$\,$M_{\odot}$, even though $\Omega _{\rm m}$ is several times greater than $\Omega _{\rm b}$ (\S \ref{s21}) and a fraction of baryon is in the gas.

\subsection{Velocity differential scale}
\label{s42}

By using Eqs. (\ref{e2}) and (\ref{e3}), from the initial power spectrum $P_{\delta}(k)$, we obtain the two-point correlation of the component $v_{\parallel}$ of the initial peculiar velocity $\mbox{\boldmath $v$} (\mbox{\boldmath $x$})$ that is parallel to the separation {\boldmath $r$}:
\begin{subequations}
\label{e15}
\begin{equation}
\label{e15a}
\langle v_{\parallel} (\mbox{\boldmath $x$}) v_{\parallel} (\mbox{\boldmath $x$} + \mbox{\boldmath $r$}) \rangle 
=   
4 \pi \dot{D}(t_{\rm in}) ^2 \int ^{\infty}_{0} \left[ j_0(kr)-\frac{2j_1(kr)}{kr} \right] P_{\delta}(k) dk.
\end{equation}
Here, $j_1(x) = \sin (x)/x^2 - \cos (x)/x$ is a spherical Bessel function. Also for the velocity vector $\mbox{\boldmath $v$} (\mbox{\boldmath $x$})$, the two-point correlation is obtained as
\begin{equation}
\label{e15b}
\langle \mbox{\boldmath $v$} (\mbox{\boldmath $x$}) \cdot \mbox{\boldmath $v$} (\mbox{\boldmath $x$} + \mbox{\boldmath $r$}) \rangle
=
4 \pi \dot{D}(t_{\rm in}) ^2 \int ^{\infty}_{0} j_0 (kr) P_{\delta}(k) dk
\end{equation}
\end{subequations}(G\'orski,\cite{Go88} see also Monin and Yaglom,\cite{MY75} p. 51). We show the profiles of $\langle v_{\parallel} (\mbox{\boldmath $x$}) v_{\parallel} (\mbox{\boldmath $x$} + \mbox{\boldmath $r$}) \rangle / \langle v_{\parallel} (\mbox{\boldmath $x$})^2 \rangle$ and $\langle \mbox{\boldmath $v$} (\mbox{\boldmath $x$}) \cdot \mbox{\boldmath $v$} (\mbox{\boldmath $x$} + \mbox{\boldmath $r$}) \rangle / \langle \vert \mbox{\boldmath $v$} (\mbox{\boldmath $x$}) \vert ^2 \rangle$ in Fig. \ref{f1}.

Equations (\ref{e6}) and (\ref{e15a}) are used to obtain the differential length scale for the $v_{\parallel}$ component:
\begin{subequations}
\label{e16}
\begin{equation}
\label{e16a}
\lambda _{v_{\parallel}}
=
\pi \left[ \frac{ 5 \int ^{\infty}_0 P_{\delta}(k) dk}{3 \int ^{\infty}_0 k^2 P_{\delta}(k) dk} \right] ^{1/2}. 
\end{equation}
Table \ref{t1} shows that the length scale $\lambda _{v_{\parallel}}$ is about 2\,Mpc and the associated mass scale $M(\lambda _{v_{\parallel}})$ is about $5 \times 10^{11}$\,$M_{\odot}$.

For reference, we use Eqs. (\ref{e6}) and (\ref{e15b}) to obtain the differential length scale for the velocity vector $\mbox{\boldmath $v$}$:
\begin{equation}
\label{e16b}
\lambda _{\mbox{\boldmath $v$}}
=
\pi \left[ \frac{ 3 \int ^{\infty}_0 P_{\delta}(k) dk}{ \int ^{\infty}_0 k^2 P_{\delta}(k) dk} \right] ^{1/2}.
\end{equation}
\end{subequations} The meaning of $\lambda$ explained in \S \ref{s31} does not hold if $\lambda$ is for a vector. Hence, $\lambda _{\mbox{\boldmath $v$}}$ is not studied here, albeit mentioned in \S \ref{s43} (see also Appendix \ref{sA}).

\begin{figure}[bp]
\begin{center}
\resizebox{6.3cm}{!}{\includegraphics*[4.9cm,14.1cm][16.cm,26.2cm]{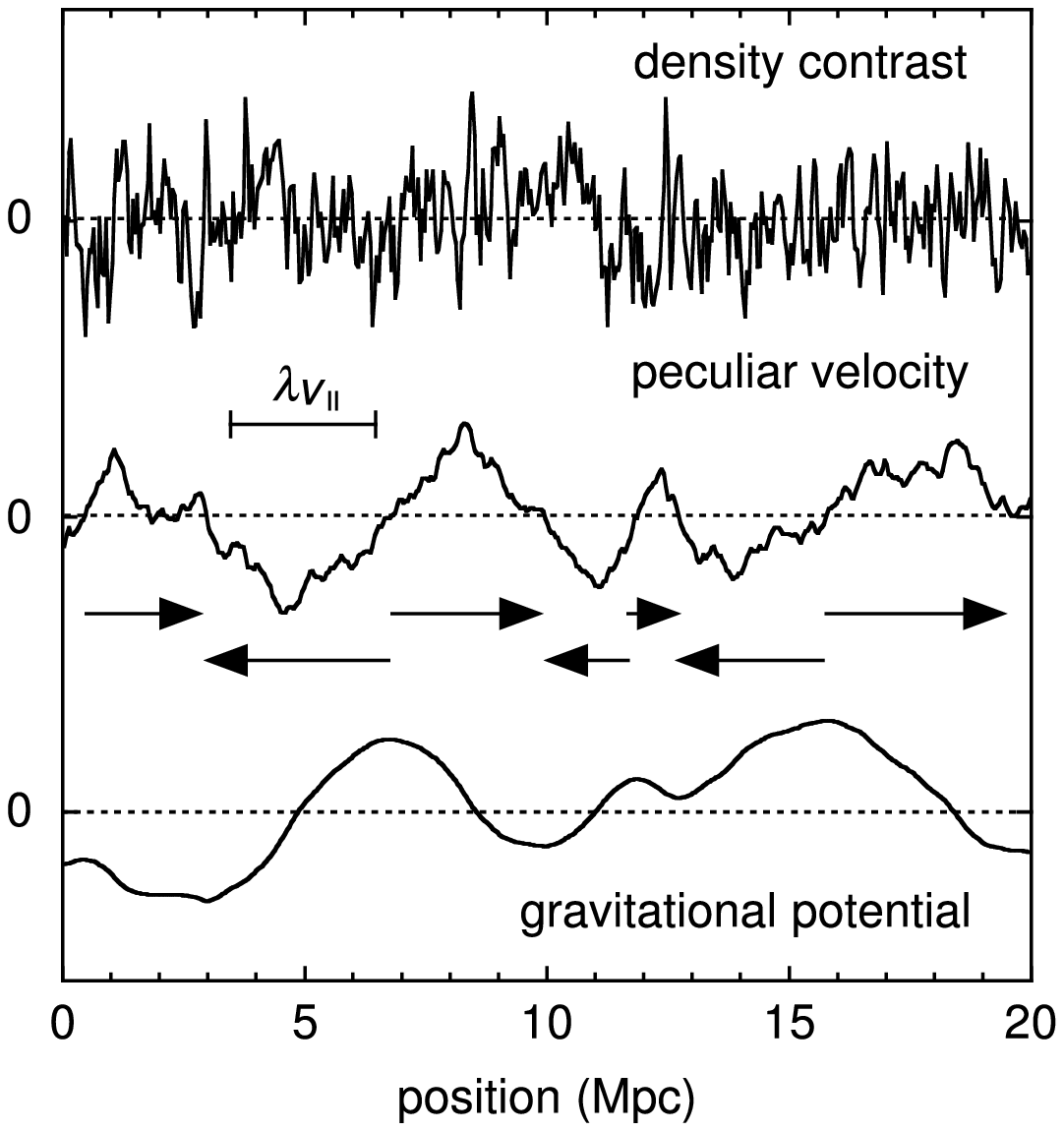}}

\caption[Fig2]{Density contrast $\delta (\mbox{\boldmath $x$})$, peculiar velocity $v_{\parallel}(\mbox{\boldmath $x$})$, and peculiar gravitational potential $\phi (\mbox{\boldmath $x$})$ at the initial time $t_{\rm in}$ along a one-dimensional cut of the space. They are in arbitrary units. The arrows indicate the clustering motions. We also show the differential length scale $\lambda _{v_{\parallel}}$.
\label{f2}}

\end{center}
\end{figure}

The mass scale $M(\lambda _{v_{\parallel}})$ serves as the typical mass for clustering toward local minima of the initial gravitational potential $\phi (\mbox{\boldmath $x$})$. This clustering is described with the initial peculiar velocity {\boldmath $v$} by the Lagrangian perturbation in Eq. (\ref{e5}). Figure \ref{f2} shows the initial fluctuations along a one-dimensional cut of the space. As indicated by arrows, there are clustering motions according to the $v_{\parallel}$ component that is parallel to the one-dimensional cut. Zeros of the $v_{\parallel}$ component are centers or boundaries of the clustering. They correspond to the local minima or maxima of $\phi (\mbox{\boldmath $x$})$ because $\mbox{\boldmath $v$}(\mbox{\boldmath $x$}) \propto -\mbox{\boldmath $\nabla$} \phi(\mbox{\boldmath $x$})$. The mean interval between those zeros is $\lambda _{v_{\parallel}}$ (\S \ref{s31}), which serves as the typical clustering radius. Hence, $M(\lambda _{v_{\parallel}}) = 4 \pi \langle \rho \rangle \lambda_{v_{\parallel}}^3/3$ serves as the typical mass that undergoes the clustering.

To be exact, the one-dimensional cut does not necessarily pass through the clustering centers in the three-dimensional space. This fact is nevertheless not serious at all, as demonstrated in Appendix \ref{sB}.

The matter elements that have clustered around a local minimum of the initial gravitational potential $\phi (\mbox{\boldmath $x$})$ interact with one another. They eventually form a self-gravitating object. This is just the prediction of the adhesion model in \S \ref{s24} (Sahni and Coles,\cite{SC95} see also Kofman et al.\cite{Ko92}). Such a self-gravitating object has the typical mass $M(\lambda _{v_{\parallel}})$.

We have studied young galaxies formed at high $z$ (\S \ref{s41}). Their halos with mass $\simeq M(L_{\delta})$ represent the matter elements that cluster to form the objects with mass $\simeq M(\lambda _{v_{\parallel}}) \gg M(L_{\delta})$ (Table \ref{t1}). On the other hand, it is widely considered that the halos of young galaxies accrete matter and merge with one another to form the larger halos of galaxies observed at $z \simeq 0$.\cite{Ba06} They are identical to the objects with mass $\simeq M(\lambda _{v_{\parallel}})$ if the accretion and mergers occur within each potential well of $\phi (\mbox{\boldmath $x$})$.

The galactic halos observed at $z \simeq 0$ had been formed almost completely at $z > 0$. Most of these halos have been accreted into the larger halos of galaxy groups or clusters but have survived as their identifiable substructures.\cite{Mo06,DL04} There also remain isolated galaxies that do not belong to groups or clusters.

The representative galaxies observed at $z \simeq 0$ are those at around the Schechter luminosity ${\cal L}_{\ast} \simeq 10^{10} {\cal L}_{\odot}$. Although the less luminous galaxies are more numerous, galaxies at around ${\cal L}_{\ast}$ dominate the luminosity integrated over all galaxies (Peebles,\cite{Pe93} p. 120), judging from the observed shape of the luminosity function. Since the mass-to-light ratio is not sensitive to the galaxy luminosity, it follows that the galaxies at around ${\cal L}_{\ast}$ share most of the mass. From the observed gravitational lensing and motions of satellite galaxies, it has been established that each isolated galaxy at ${\cal L}_{\ast}$ has the total halo mass of $(1$--$2) \times 10^{12}$\,$M_{\odot}$.\cite{Za93,Br96,Hoek05} The same halo mass is obtained for galaxies at ${\cal L}_{\ast}$ in groups and clusters, except for those in the cluster core where the tidal field is too strong.\cite{Ma06,Na09} They are close to $M(\lambda _{v_{\parallel}}) \simeq 5 \times 10^{11}$\,$M_{\odot}$.

\subsection{Velocity integral scale}
\label{s43}

Equations (\ref{e7}) and (\ref{e15}) are used to obtain the integral length scales for the initial peculiar velocity $\mbox{\boldmath $v$} (\mbox{\boldmath $x$})$ from the initial power spectrum $P_{\delta}(k)$:
\begin{subequations}
\label{e17}
\begin{align}
&
L_{v_{\parallel}}
=     
0, \\
&
L_{\mbox{\boldmath $v$}}
=     
\frac{\pi \int ^{\infty}_0 k^{-1} \, P_{\delta}(k) dk}{2 \int ^{\infty}_0 P_{\delta}(k) dk}. 
\label{e17a} 
\end{align}
\end{subequations}The length scale $L_{v_{\parallel}}$ is always zero because $\langle v_{\parallel} (\mbox{\boldmath $x$}) v_{\parallel} (\mbox{\boldmath $x$} + \mbox{\boldmath $r$}) \rangle$ changes its sign (Fig. \ref{f1}), while the length scale $L_{\mbox{\boldmath $v$}}$ is about 90\,Mpc (Table \ref{t1}).\footnote{
The mass scale $M(L_{\mbox{\boldmath $v$}}) = 4 \pi \langle \rho \rangle L_{\mbox{\boldmath $v$}}^3/3$ is not considered here because the corresponding structures are not self-gravitating even at $z \simeq 0$ (see below).}

Table \ref{t1} shows $L_{\mbox{\boldmath $v$}} \gg \lambda_{\mbox{\boldmath $v$}} \simeq 2$ Mpc. Beneath the small-scale fluctuations that dominate the differential length scale $\lambda_{\mbox{\boldmath $v$}}$ (see Fig. \ref{f2}), there is hidden the large-scale correlation that dominates the integral length scale $L_{\mbox{\boldmath $v$}}$ (see Fig. \ref{f1}).

The length scale $L_{\mbox{\boldmath $v$}}$ serves as the typical size for the largest structures of the Universe such as those in the galaxy distribution. Even at $z \simeq 0$, their formation is ongoing and in the quasi-linear stage. The Lagrangian perturbation in Eq.~(\ref{e5}) implies that the formation is due to large-scale correlated motions in the initial velocity field $\mbox{\boldmath $v$} (\mbox{\boldmath $x$})$.\cite{SC95,Bo96} As shown in Fig. \ref{f1}, the correlation of $\mbox{\boldmath $v$} (\mbox{\boldmath $x$})$ is significant up to the scale $r \simeq L_{\mbox{\boldmath $v$}}$ (\S \ref{s31}). The structure formation is accordingly significant. Above the scale $r \simeq L_{\mbox{\boldmath $v$}}$, the correlation is not significant, and hence the structure formation is not significant. Thus, on length scales for the significant structure formation in the quasi-linear stage, there is an upper limit $L_{\mbox{\boldmath $v$}}$.

We underline that the above discussion is statistical. Some structures could be formed far above the scale $r \simeq L_{\mbox{\boldmath $v$}}$, where $\langle \mbox{\boldmath $v$} (\mbox{\boldmath $x$}) \cdot \mbox{\boldmath $v$} (\mbox{\boldmath $x$} + \mbox{\boldmath $r$}) \rangle / \langle \vert \mbox{\boldmath $v$}(\mbox{\boldmath $x$}) \vert ^2 \rangle$ is not significant but is not yet absent (Fig. \ref{f1}). The initial velocity field $\mbox{\boldmath $v$} (\mbox{\boldmath $x$})$ could be correlated over very large scales in some exceptional regions, which could form very large structures. However, such exceptional structures are not statistically important.

The galaxy distribution observed at $z \simeq 0$ has large-scale structures such as filaments, walls, and voids. Statistical signatures for the existence of the structures are significant up to 50--100\,Mpc, above which the galaxy distribution is almost uniform.\cite{Ku01,Jo04,Hogg05} This length scale for transition to uniformity is close to $L_{\mbox{\boldmath $v$}} \simeq 90$\,Mpc. Although structures much larger than $L_{\mbox{\boldmath $v$}}$ are known, i.e., `great walls',\cite{Go05} they are exceptional and are not significant to the statistics.\cite{Ku01}

The observed galaxy distribution is also used to reconstruct the peculiar velocity field at $z \simeq 0$,\cite{De94,Er06} where we see large-scale correlated motions from voids toward filaments and walls. The length scales of these motions, albeit not explicitly given in the literature, are up to about $L_{\mbox{\boldmath $v$}} \simeq 90$\,Mpc.

According to the adhesion model in \S \ref{s24}, the structure formation has two stages (Sahni and Coles,\cite{SC95} see also Kofman et al.\cite{Ko92}). The first stage is clustering of matter elements toward local minima of the initial gravitational potential $\phi(\mbox{\boldmath $x$})$ and the resultant formation of self-gravitating objects. The second stage is clustering of these objects due to a large-scale correlation in the initial gravitational acceleration $-\mbox{\boldmath $\nabla$} \phi(\mbox{\boldmath $x$})$. Up to some characteristic scale of the correlation, the structure formation is significant. By characterizing the initial peculiar velocity $\mbox{\boldmath $v$}(\mbox{\boldmath $x$}) \propto -\mbox{\boldmath $\nabla$} \phi(\mbox{\boldmath $x$})$ with the new scales $\lambda _{v_{\parallel}}$ and $L_{\mbox{\boldmath $v$}}$, we have found that, while formed in the first stage are halos of galaxies with mass $\simeq M(\lambda _{v_{\parallel}})$ (\S \ref{s42}), formed in the second stage are the largest structures in the galaxy distribution with size $\simeq L_{\mbox{\boldmath $v$}}$.

\section{Conclusion}
\label{s5}

For the initial fields of the density contrast $\delta$ and peculiar velocity {\boldmath $v$}, we have calculated new statistical measures: the differential length scale $\lambda$, the integral length scale $L$, and the associated mass scales $M(\lambda)$ and $M(L)$. Through the first-order Eulerian or Lagrangian perturbation, these scales are related to scales for the structure formation. We have found that they explain the scales observed for galaxies and their distribution.

The mass scale $M(L_{\delta})$ at $> 2 \times 10^6$\,$M_{\odot}$ serves as the typical mass for overdense regions around high density peaks of the initial fluctuations, which explains the typical mass for halos of young galaxies observed at $z \gtrsim 5$.

The mass scale $M(\lambda _{v_{\parallel}}) \simeq 5 \times 10^{11}$\,$M_{\odot}$ serves as the typical mass for clustering toward local minima of the initial gravitational potential, which explains the typical mass for halos of galaxies observed at $z \simeq 0$.

The length scale $L_{\mbox{\boldmath $v$}} \simeq 90$\,Mpc serves as the typical size for the largest structures formed in the Universe, which explains the length scale for transition to uniformity in the galaxy distribution observed at $z \simeq 0$.

Therefore, these observed scales are fossil imprints of the characteristic scales of the initial fields. While the halos of young galaxies with mass $\simeq M(L_{\delta})$ are extinct at $z \simeq 0$, the halos of galaxies with mass $\simeq M(\lambda _{v_{\parallel}})$ and the largest structures with size $\simeq L_{\mbox{\boldmath $v$}}$ are existent and observable at $z \simeq 0$.


\appendix
\section{Other Characteristic Scales of the Peculiar Velocity}
\label{sA}

The differential and integral scales also exist for the $v_{\perp}$ component of the initial peculiar velocity that is perpendicular to the separation {\boldmath $r$}. Although this $v_{\perp}$ component is not directly relevant to the structure formation studied here, we summarize its scales for reference.

By using Eqs.~(\ref{e2}) and (\ref{e3}), we obtain the two-point correlation of the $v_{\perp}$ component\cite{Go88}
\begin{equation}
\label{eA1}
\langle v_{\perp} (\mbox{\boldmath $x$}) v_{\perp} (\mbox{\boldmath $x$} + \mbox{\boldmath $r$}) \rangle
=
4 \pi \dot{D}(t_{\rm in}) ^2 \int ^{\infty}_{0} \frac{j_1(kr)}{kr} P_{\delta}(k) dk .
\end{equation}
The profile of $\langle v_{\perp} (\mbox{\boldmath $x$}) v_{\perp} (\mbox{\boldmath $x$} + \mbox{\boldmath $r$}) \rangle / \langle v_{\perp} (\mbox{\boldmath $x$})^2 \rangle$ is shown in Fig. \ref{f1}. Equations (\ref{e6}), (\ref{e7}), and (\ref{eA1}) are used to obtain the length scales
\begin{align}
& \label{eA2}
\lambda _{v_{\perp}}
=
\pi \left[ \frac{ 5 \int ^{\infty}_0 P_{\delta}(k) dk}{ \int ^{\infty}_0 k^2 P_{\delta}(k) dk} \right] ^{1/2}, \\
& \label{eA3}
L_{v_{\perp}}
=     
\frac{3\pi \int ^{\infty}_0 k^{-1} \, P_{\delta}(k) dk}{4 \int ^{\infty}_0 P_{\delta}(k) dk}.
\end{align}
Table \ref{t1} lists the numerical values of these length scales and of the associated mass scales.

\section{Clustering Mass Obtained from the Gravitational Potential}
\label{sB}

We have discussed the mass scale $M(\lambda _{v_{\parallel}})$ as the typical mass for clustering toward local minima of the initial gravitational potential $\phi (\mbox{\boldmath $x$})$, by ignoring the fact that a one-dimensional cut does not necessarily pass through the local minima in the three-dimensional space (\S \ref{s42}). Here is a more exact discussion that leads to almost the same result.

We calculate the number density of the clustering centers, i.e., number density $n_{{\rm min},\phi}$ of local minima of the initial gravitational potential $\phi(\mbox{\boldmath $x$})$. For a three-dimensional random scalar field that is homogeneous, isotropic, and Gaussian as in the case of $\phi(\mbox{\boldmath $x$})$, the number density $n_{\rm max}$ of the local maxima was calculated by Bardeen et al.\cite{Ba86} To their Eqs. (4$\cdot$6) and (4$\cdot$11), we substitute the power spectrum of $\phi(\mbox{\boldmath $x$})$ obtained from Eqs. (\ref{e1}), (\ref{e2}), and (\ref{e4}), which is proportional to $P_{\delta}(k)/k^4$. Then,
\begin{equation}
\label{eqB1}
n_{{\rm min},\phi}
=
n_{{\rm max},\phi}
=
\frac{29-6\sqrt{6}}{40\sqrt{5} \pi^2} r_{\phi}^{-3}
\quad
\mbox{with}
\quad
r_{\phi}
=
\left[ \frac{ 3 \int ^{\infty}_0 P_{\delta}(k) dk}{ \int ^{\infty}_0 k^2 P_{\delta}(k) dk} \right] ^{1/2}.
\end{equation}
We have $r_{\phi} \propto \lambda _{v_{\parallel}}$ (Eq. (\ref{e16a})), although $r_{\phi}$ does not have any particular meaning. The number density $n_{{\rm min},\phi}$ leads to the typical mass $\langle \rho \rangle / n_{{\rm min},\phi}$ for clustering toward local minima of the initial gravitational potential $\phi(\mbox{\boldmath $x$})$, which is only 15\% greater than $M(\lambda _{v_{\parallel}}) = 4 \pi \langle \rho \rangle \lambda _{v_{\parallel}}^3/3$. In fact, our cosmological parameters yield $\langle \rho \rangle / n_{{\rm min},\phi} = 5.7^{-1.4}_{+1.9} \times 10^{11} \,M_{\odot}$, while $M(\lambda _{v_{\parallel}}) = 5.0^{-1.3}_{+1.7} \times 10^{11}\,M_{\odot}$.

\end{document}